\documentclass[sigconf,screen]{acmart}

\usepackage{multirow}
\usepackage{graphicx}
\usepackage{subcaption}
\usepackage{xspace}
\usepackage{algorithm}
\usepackage[noend]{algpseudocode}
\usepackage[normalem]{ulem}
\useunder{\uline}{\ul}{}
\usepackage{ragged2e}
\usepackage{array}
\usepackage{amsmath}
\usepackage{xcolor}
\usepackage{balance}
\usepackage{booktabs}
\usepackage[flushleft]{threeparttable}
\usepackage{enumitem}
\usepackage{pifont}
\usepackage{listings}

\newcolumntype{P}[1]{>{\centering\arraybackslash}p{#1}}

\newcommand{\benchmarkName}[0]{\textsc{Sphinx}}
\newcommand{\companyX}[0]{Tencent}

\newcommand{\appX}[0]{WeChat}
\newcommand{\taskOpen}[0]{244}
\newcommand{\taskWeChat}[0]{214}

\newcommand{\appTotal}[0]{100}
\newcommand{\categoryTotal}[0]{17}

\newcommand{\lmNum}[0]{8}
\definecolor{DarkGreen}{RGB}{1,100,32}
\definecolor{DarkRed}{RGB}{158,19,22}
\newcommand{\cmark}{\textcolor{DarkGreen}{\ding{51}}}%
\newcommand{\xmark}{\textcolor{DarkRed}{\ding{55}}}%

\newcommand{\Answer}[2]{\noindent \textbf{Lessons Learned from #1:} #2}
\usepackage{tikz}
\definecolor{customblue}{RGB}{192, 208, 235}
\definecolor{customdarkblue}{RGB}{68, 114, 196}

\newcommand{\custombluecircled}[1]{%
  \tikz[baseline=(char.base)]{
    \node[shape=circle, color=customdarkblue, draw, fill=customblue, text=black,  font=\sffamily, inner sep=1pt] (char) {#1};
  }%
}

\begin{document}

\title[]{Beyond Pass or Fail: Multi-Dimensional Benchmarking of Foundation Models for Goal-based Mobile UI Navigation}

\author{Dezhi Ran}
\authornote{Project leaders; Equal contribution.}
\affiliation{%
  \department{Key Lab of HCST (PKU), MOE; SCS}
  \institution{Peking University}
  \city{Beijing}
  \country{China}
}
\email{dezhiran@pku.edu.cn}

\author{Mengzhou Wu}
\authornotemark[1]
\affiliation{%
  \institution{School of EECS, Peking University}
  \city{Beijing}
  \country{China}
}
\email{wmz@stu.pku.edu.cn}

\author{Hao Yu}
\affiliation{%
  \institution{School of Software and Microelectronics, Peking University}
  \city{Beijing}
  \country{China}
}
\email{yh0315@pku.edu.cn}

\author{Yuetong Li}
\affiliation{%
  \institution{The University of Chicago}
  \city{Chicago}
  \country{USA}
}
\email{yuetong@uchicago.edu}

\author{Jun Ren}
\affiliation{%
  \institution{University of Texas at Dallas}
  \city{Dallas}
  \country{USA}
}
\email{jxr210020@utdallas.edu}

\author{Yuan Cao}
\affiliation{%
  \institution{School of EECS, Peking University}
  \city{Beijing}
  \country{China}
}
\email{cao_yuan21@stu.pku.edu.cn}

\author{Xia Zeng}
\affiliation{%
  \institution{Tencent Inc.}
  \city{Shenzhen}
  \country{China}
}
\email{xiazeng@tencent.com}

\author{Haochuan Lu}
\affiliation{%
  \institution{Tencent Inc.}
  \city{Shenzhen}
  \country{China}
}
\email{hudsonhclu@tencent.com}

\author{Zexin Xu}
\affiliation{%
  \institution{University of Texas at Dallas}
  \city{Dallas}
  \country{USA}
}
\email{Zexin.xu@utdallas.edu}

\author{Mengqian Xu}
\affiliation{%
  \institution{East China Normal University}
  \city{Shanghai}
  \country{China}
}
\email{xmq@stu.ecnu.edu.cn}

\author{Ting Su}
\affiliation{%
  \institution{East China Normal University}
  \city{Shanghai}
  \country{China}
}
\email{tsu@sei.ecnu.edu.cn}

\author{Liangchao Yao}
\affiliation{%
  \institution{Tencent Inc.}
  \city{Shenzhen}
  \country{China}
}
\email{clarkyao@tencent.com}

\author{Ting Xiong}
\affiliation{%
  \institution{Tencent Inc.}
  \city{Shenzhen}
  \country{China}
}
\email{candyxiong@tencent.com}

\author{Wei Yang}
\affiliation{%
  \institution{University of Texas at Dallas}
  \city{Dallas}
  \country{USA}
}
\email{wei.yang@utdallas.edu}

\author{Yuetang Deng}
\affiliation{%
  \institution{Tencent Inc.}
  \city{Shenzhen}
  \country{China}
}
\email{yuetangdeng@tencent.com}

\author{Assaf Marron}
\affiliation{%
  \department{Dept. of Computer Science and Applied Mathematics}
  \institution{Weizmann Institute of Science}
  \city{Rehovot}
  \country{Israel}
}
\email{Assaf.Marron@weizmann.ac.il}

\author{David Harel}
\affiliation{%
  \department{Dept. of Computer Science and Applied Mathematics}
  \institution{Weizmann Institute of Science}
  \city{Rehovot}
  \country{Israel}
}
\email{david.harel@weizmann.ac.il}

\author{Tao Xie}
\authornote{Tao Xie is the corresponding author.}
\affiliation{%
  \department{Key Lab of HCST (PKU), MOE; SCS}
  \institution{Peking University}
  \city{Beijing}
  \country{China}
}
\email{taoxie@pku.edu.cn}

\renewcommand{\shortauthors}{Dezhi Ran et al.}

\begin{abstract}
Recent advances of foundation models (\textit{FMs}) have made navigating mobile applications (\textit{apps}) based on high-level goal instructions within reach, with significant industrial applications such as UI testing. While existing benchmarks evaluate FM-based UI navigation using the binary pass/fail metric, they have two major limitations: they cannot reflect the complex nature of mobile UI navigation where FMs may fail for various reasons (e.g., misunderstanding instructions and failed planning), and they lack industrial relevance due to oversimplified tasks that poorly represent real-world scenarios.
To address the preceding limitations, we propose \benchmarkName{}, a comprehensive benchmark for multi-dimensional evaluation of FMs in industrial settings of UI navigation. \benchmarkName{} introduces a specialized toolkit that evaluates five essential FM capabilities, providing detailed insights into failure modes such as insufficient app knowledge or planning issues. Using both popular Google Play applications and WeChat's internal UI test cases, we evaluate \lmNum{} FMs with 20 different configurations. Our results show that existing FMs universally struggle with goal-based testing tasks, primarily due to insufficient UI-specific capabilities. We summarize seven lessons learned from benchmarking FMs with \benchmarkName{}, providing clear directions for improving FM-based mobile UI navigation.

\end{abstract}


\maketitle

\section{Introduction}\label{sec::intro}

\begin{figure*}[t]
        \centering
\includegraphics[width=0.95\linewidth]{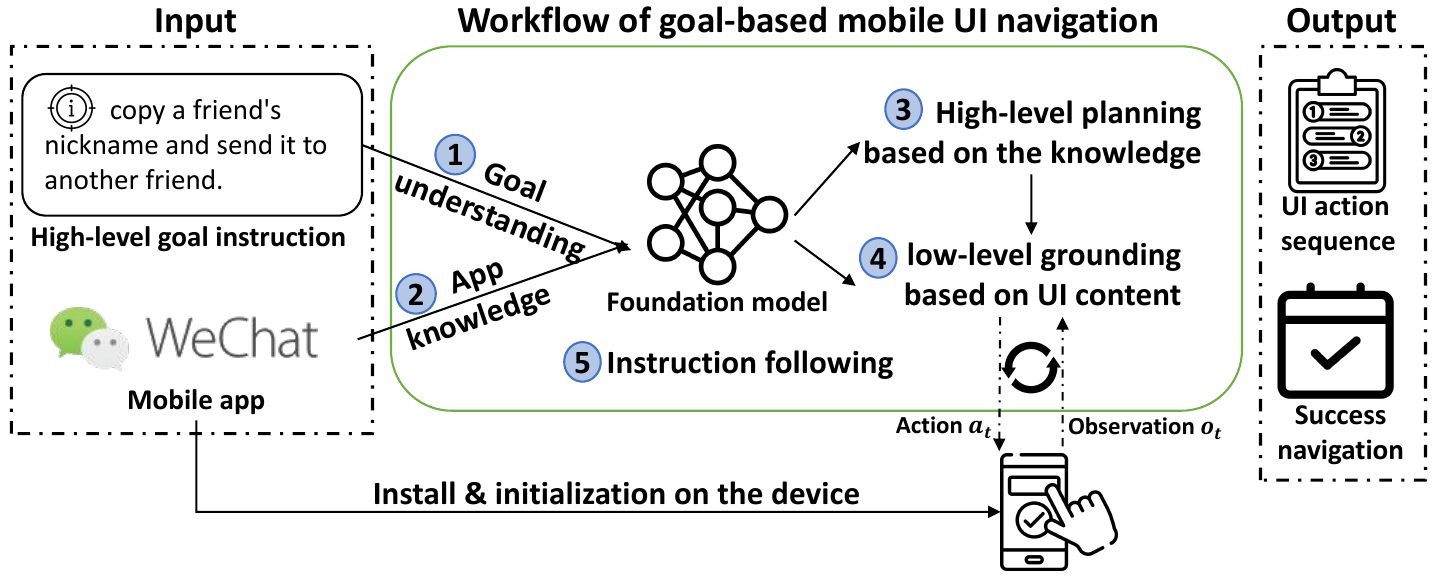}
        \caption{An example of goal-based mobile UI navigation on WeChat with a foundation model. 
        }
        \label{fig::agent_workflow}
    \end{figure*}
    
To enhance app accessibility~\cite{naftali2014accessibility, griffin2017survey, yan2019current, chen2020unblind, ran2024elderly} and reduce quality-assurance costs~\cite{wei2016taming, linares2017developers, dobslaw2019estimating,ran2024guardian}, automated navigation through mobile application (\textit{app}) UIs based on high-level goal instructions~\cite{li2020mapping}, denoted as \textit{goal-based UI navigation}, is an increasingly desirable solution~\cite{yao2022webshop,wen2024autodroid,ran2024guardian} for industry.
As shown in Figure~\ref{fig::agent_workflow}, goal-based UI navigation
explores the target app to find a sequence of UI actions (i.e., UI events) to achieve a given goal without bothering users to operate their devices. 
Despite the usefulness, implementing goal-based UI navigation requires five key capabilities: understanding the goal instruction (\custombluecircled{1}, \textit{goal understanding}), extracting app knowledge (\custombluecircled{2}, \textit{app knowledge}), high-level planning (\custombluecircled{3}, \textit{planning}), grounding the plan to the UI content (\custombluecircled{4}, \textit{grounding}), and following specific instructions during navigation (\custombluecircled{5}, \textit{instruction following}), conducting goal-based UI navigation has been a long-standing open challenge~\cite{thummalapenta2012automating}.

Recent advances of \textit{foundation models}~\cite{bommasani2021opportunities} (in short as \textit{FMs}) make goal-based UI navigation within reach.
FMs are large-scale, pre-trained models~\cite{bommasani2021opportunities} that can be adapted to a wide range of downstream tasks~\cite{brown2020language}.
Trained on massive datasets and possessing a broad understanding of various domains~\cite{touvron2023llama, openai2023gpt4}, FMs are shown to be capable of understanding and following user instructions in various question-answering tasks~\cite{lu2023multi, zaib2022conversational, singhal2025toward}, understanding UI contents in tasks of summarizing screen contents~\cite{wang2021screen2words}, and conducting planning based on high-level goals~\cite{song2023llmplanner}.
The success of FMs on individual tasks makes them promising to satisfy the multiple requirements depicted in Figure~\ref{fig::agent_workflow} for implementing approaches of goal-based UI navigation, and exploring their potentials in UI navigation is becoming a hot research and industrial field~\cite{yao2022react, wen2023droidbot, wen2023empowering, yang2023appagent, liu2024make, wen2024autodroid,ran2024guardian, anthropic_computer_use}.

Despite the great potentials of FMs, recent benchmarks~\cite{zhou2023webarena,wen2024autodroid,ran2024guardian,xing2024androidarena} on UI navigation are used to evaluate FMs' end-to-end effectiveness and all of the benchmarks demonstrate that FMs  exhibit low effectiveness.
DroidTask~\cite{wen2024autodroid} provides 158 Android UI navigation tasks from 13 open-source apps and compares the ground-truth UI trajectory to determine whether an FM succeeds on the task or not. The state-of-the-art FM namely GPT-4 succeeds on 36\% of all tasks.
FestiVal~\cite{ran2024guardian} extends the evaluation scope from open-source apps to industrial apps consisting of 70 UI navigation tasks from popular industrial apps, where GPT-3.5 with agent design of Reflexion~\cite{yao2022react} succeeds on only 19\% of the tasks.

Despite their usefulness in demonstrating low effectiveness of FMs and inspiring agent designs to improve FM-based UI navigation~\cite{yang2023appagent,wen2024autodroid,ran2024guardian}, existing benchmarks fail to satisfy two major requirements for fully understanding and improving FMs in real-world UI navigation, especially in industrial scenarios.
First, multi-dimensional evaluation is lacking, as existing benchmarks primarily focus on end-to-end success rates without assessing fine-grained capabilities required for UI navigation. This narrow scope provides no support for a comprehensive quantification of LLM agent abilities, limits the understanding of their specific weaknesses, and hinders the identification of failure points necessary to guide targeted improvements.
Second, industrial relevance is limited, as the tasks collected by existing benchmarks are often simplified, failing to reflect the complexity and diversity of real-world industrial applications.
As a result, these benchmarks may lack generalizability and fail to address the challenges encountered in real-world environments, ultimately hindering their applicability in industrial settings.

To bridge the preceding gaps, in this paper, we present \textbf{\benchmarkName{}}, the first multi-dimensional benchmark of FMs for goal-based mobile UI navigation in an industrial setting. \benchmarkName{} aims to bridge these gaps through the following two key designs.

\noindent\textbf{Comprehensive toolkits for multi-dimensional evaluation.}
In addition to evaluating the end-to-end effectiveness, \benchmarkName{} evaluates the five capabilities depicted in Figure~\ref{fig::agent_workflow} with specialized evaluation tasks.
By systematically isolating and analyzing each capability, \benchmarkName{} provides a comprehensive framework for understanding the performance of FMs in goal-based UI navigation.
This evaluation goes beyond the simple pass/fail metric, enabling researchers to identify specific weaknesses and gain deeper insights into the root causes of failures, thereby highlighting future directions to enhance the end-to-end effectiveness of FMs in UI navigation.

\noindent\textbf{Representative-task collection.}
To complement existing benchmarks where UI navigation tasks usually come from a limited selection of a few open-source apps~\cite{wen2024autodroid} and without testing tasks,
we design \benchmarkName{} to focus on tasks drawn from real industry practices with two major approaches of task collection.
First, we utilize UI test cases used in the daily quality-assurance processes of \appX{}, a highly popular industrial app with over \textit{one billion} monthly active users.
These testing tasks are collected and annotated by three quality assurance (QA) engineers of WeChat and represent the engineers' expectations of goal-based UI navigation.
Second, we collect \taskOpen{} user tasks on \appTotal{} popular industrial apps from \categoryTotal{} categories to cover popular UI navigation tasks.
These tasks represent  common users' expectations of goal-based UI navigation.

With the proposed \benchmarkName{}, we conduct comprehensive evaluations with \lmNum{} models, including both state-of-the-art proprietary models (e.g., GPT-4o), popular open-source models (e.g., Llama3), as well as popular UI navigation agents ReAct~\cite{yao2022react} and AppAgent\cite{yang2023appagent}.

Our evaluations show that there is still a long way before FM-based UI-navigation. Consistent with benchmarking results on other real-world problems~\cite{zhou2023webarena,swe-bench,mialon2023gaiabenchmarkgeneralai}, all FMs achieve low effectiveness on \benchmarkName{}. The performance further worsens on testing tasks, where none of the FMs succeeds in solving a single task.
Furthermore, our results of multi-dimensional evaluation highlight three key findings as well as seven lessons learned to improve FMs for mobile UI-navigation:

\noindent\textbf{Vision modality lags significantly behind text modality in mobile UI navigation.} Our evaluation results demonstrate that the vision modality exhibits substantially lower performance compared to the text modality, even when incorporating state-of-the-art techniques such as Set of Marks (SoM)~\cite{yang2023setofmark}. These findings strongly suggest that the design of FM-based UI navigation agents should give priority to  text modality being fed to FMs, while leveraging vision input primarily as a supplementary information source.

\begin{figure}[t]
        \centering
\includegraphics[width=\linewidth]{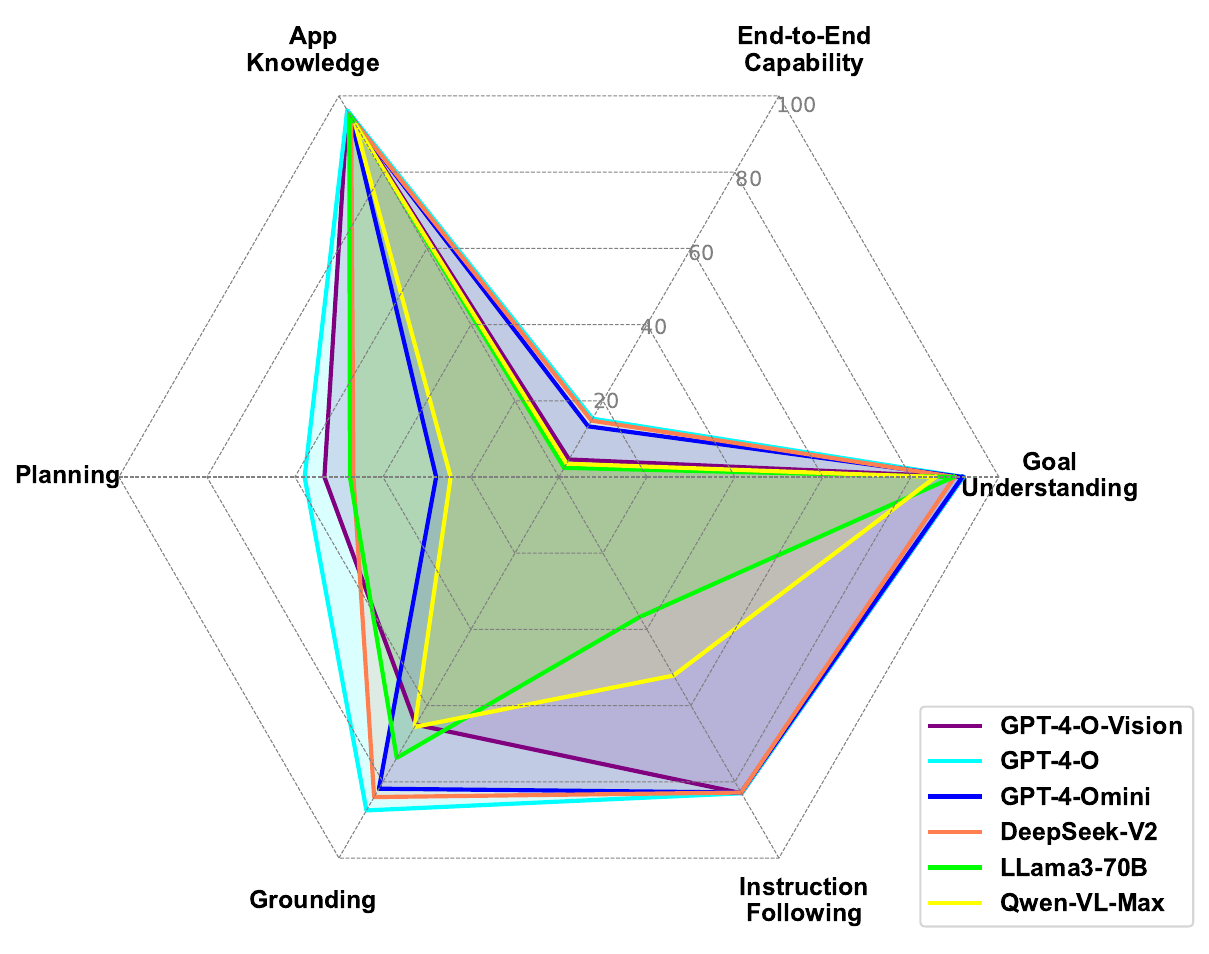}
        \caption{Visualization of popular FMs' five key capabilities required for mobile UI navigation and end-to-end effectiveness on \benchmarkName{}. UI-specific capabilities are the bottlenecks for FM-based mobile UI navigation.}
        \label{fig::radar}
    \end{figure}

\noindent\textbf{UI-specific capabilities manifest as critical bottlenecks of FMs.}
Beyond pass or fail, the multi-dimensional evaluation reveals the key shortcomings of FMs  resulting the low end-to-end effectiveness, i.e., lacking UI-specific capabilities.
As shown in Figure~\ref{fig::radar},
while the existing FMs are experts at general language processing such as goal understanding and possess rich knowledge about how to complete the given task, they fall short in transforming the knowledge into actionable planning relevant to the given UI contents and following the given instructions~\cite{ouyang2022training,sun2023evaluating}. 

\noindent \textbf{Limitations of FMs invalidate  sophisticated agent designs.}
Through an overall analysis and case study of AppAgent under different FMs, we find  that the agent's performance is constrained by the underlying FM's limitations revealed by \benchmarkName{}, highlighting the need for training UI-specific FMs to better support UI navigation tasks.

In summary, this paper makes the following major contributions: 
\begin{itemize}
    \item We propose a novel multi-dimensional benchmark, \benchmarkName{},  comprising five distinct tasks, designed to comprehensively evaluate the core capabilities of FMs required for mobile UI navigation.
    
    \item We implement an automated and ready-to-use and benchmarking suite, along with representative tasks collected from highly popular mobile apps. Both \benchmarkName{} and its scripts of collecting user tasks are publicly available~\cite{sphinx-website}, enabling easy adoption and customization for future research.
    
    \item Extensive evaluations with \lmNum{} popular FMs on \benchmarkName{} and seven lessons learned pinpoint the importance of multi-dimensional evaluation and future directions to improve FMs for UI navigation.
\end{itemize}

\section{Related Work}\label{sec::related}
\begin{table*}[t]
	\centering
    \caption{Comparison of \benchmarkName{} to existing UI navigation benchmarks. 
    }
    \label{table::dataset-comparison}
        \begin{threeparttable}
        \resizebox{\linewidth}{!}{
        \begin{tabular}{llccccccc}
            \toprule
            \multirow{2}{*}{Benchmark} & \multirow{2}{*}{Platform} & \multicolumn{1}{c}{\# Apps or} &\# Task & \multicolumn{1}{c}{Fine-grained} & Popular & \multicolumn{1}{c}{Multi-dim.} & Testing & Observation \\
            &   & websites & length  & annotation & industry app & evaluation & task & \multicolumn{1}{c}{format} \\
            \midrule
            OS-World~\cite{OSWorld} & desktop  &  N/A   & N/A & \xmark & \cmark &  \xmark   & \xmark & Screen \\
            WebShop~\cite{yao2022webshop} & web  &  1   & 7.3 & \xmark & \xmark & \xmark   & \xmark & DOM\\
            Mind2Web~\cite{deng2024mind2web}  & web   &  137   & 7.3 & \cmark & \cmark & \xmark   & \xmark & DOM, screen \\
            WebArena~\cite{zhou2023webarena} & web  &  6   & N/A & \xmark & \cmark & \xmark   & \xmark & DOM \\
            VisualArena~\cite{koh2024visualwebarena} & web   &  6   & N/A & \xmark & \cmark & \xmark   & \xmark & Screen\\
            AndroidArena~\cite{xing2024androidarena} & Android & 13 & 6.6 & \xmark & \xmark & \xmark & \xmark & A11y Text, screen\\
            DroidTask~\cite{wen2024autodroid} & Android & 13 & 4.8 & \cmark & \xmark & \xmark & \xmark & A11y Text, screen \\
            \midrule
            \textbf{\benchmarkName{}}  & Android &  \appTotal{}  & \textbf{8.1} & \cmark & \cmark & \cmark   & \cmark & A11y Text, screen \\
            \bottomrule
        \end{tabular}
        }
    \end{threeparttable}

\end{table*}
\subsection{UI Navigation Benchmarks}
Multiple UI navigation benchmarks on UI navigation have been proposed~\cite{yao2022webshop,zhou2023webarena,koh2024visualwebarena, OSWorld, deng2024mind2web, xing2024understanding, wen2024autodroid}, as shown in Table~\ref{table::dataset-comparison}.
WebShop~\cite{yao2022webshop} designs a synthetic e-commerce website to evaluate the capability of LLMs to understand user instructions and conduct planning correctly grounded with web content. 
Recently, WebArena~\cite{zhou2023webarena} and OS-World~\cite{OSWorld} benchmark the functional correctness of foundation models on real-world web and desktop navigation tasks.
VisualWebArena~\cite{koh2024visualwebarena} extends WebArena by providing vision modality observations.
DroidTask~\cite{wen2024autodroid} focuses on UI navigation tasks specific to the Android platform, while AndroidArena~\cite{xing2024understanding} evaluates five distinct capabilities through various metrics in end-to-end UI navigation.

Despite their usefulness of demonstrating the ineffectiveness of existing models in UI navigation, none of the existing benchmarks provide a multi-dimensional evaluation as \benchmarkName{} does.
Moreover, none of the existing benchmarks collect professional UI navigation testing tasks, i.e., real test generation tasks as benchmarking subjects.

\subsection{Foundation Models for UI Navigation}
Existing work on training and exploiting foundation models for UI navigation can be classified into three categories.
First, existing work utilizes self-supervised~\cite{bai2021uibert,li2021screen2vec,lee2023pix2struct} and supervised
methods~\cite{chen2020unblind,rawles2024androidinthewild} to train UI large models for UI understanding~\cite{bai2021uibert,li2021screen2vec}, widget captioning~\cite{lee2023pix2struct} and widget retrieval~\cite{bai2021uibert}.
Second, existing work utilizes foundation models to assist or conduct various UI navigation tasks.
Liu et al.~\cite{liu2022fill} use ChatGPT to synthesize textual inputs during the UI navigation.
Feng et al.~\cite{feng2024prompting} use ChatGPT to reproduce bug reports on the given mobile apps by grounding step-wise descriptions to UI elements.
Liu et al.~\cite{liu2024make} use ChatGPT to replace traditional automated UI testing by transforming automated UI testing~\cite{yang13:grey,su2017guided,ran2022automated,ran2023badge} into a question-answering problem~\cite{antol2015vqa}.
Third, a growing body of work focuses on goal-based mobile UI navigation~\cite{li2020mapping,wen2023droidbot,wen2023empowering,yang2023appagent,ran2024guardian}. Droidbot~\cite{wen2023droidbot,wen2023empowering} and AppAgent~\cite{yang2023appagent} use ChatGPT and GPT-4V to automated goal-based UI navigation.
Guardian~\cite{ran2024guardian} designs runtime system support for improving LLM-based UI navigation.
Given the growing trend of developing foundation models for goal-based mobile UI navigation, \benchmarkName{} can benefit the community by providing a comprehensive benchmark.

\section{\benchmarkName{} Benchmark}\label{sec::benchmark}

\subsection{Overview of \benchmarkName{}}
In this section, we first present the overview of \benchmarkName{}, a multi-dimensional benchmark for evaluating different capabilities of foundation models required for mobile UI navigation.

\noindent\textbf{Benchmarking dimensions.}
\benchmarkName{} evaluates five critical dimensions essential for effective mobile UI navigation, alongside assessing overall end-to-end performance. These dimensions are: (1) Goal understanding (as shown in Figure~\ref{fig::agent_workflow}, \custombluecircled{1}), evaluating the accuracy of the model's interpretation of user intents; (2) App knowledge proficiency (\custombluecircled{2}), assessing the model's understanding of app-specific information; (3) Planning capability (\custombluecircled{3}), gauging the model's capacity to infer and execute steps required for task completion; (4) Grounding capability (\custombluecircled{4}), ensuring precise mapping of low-level instructions to corresponding UI elements; and (5) Instruction following (\custombluecircled{5}), measuring the model's accuracy in executing user-provided directives. Toolkits used to evaluate these dimensions are detailed in Section~\ref{subsec::eval_dims}, while the task collection process for each dimension is described in Section~\ref{subsec::task_collection}.

\noindent\textbf{Automated benchmarking interface.}
We provide a ready-to-use benchmarking interface to facilitate future research with \benchmarkName{}.
We provide four observation modes to support the evaluation of foundational models with various input modalities (Section~\ref{subsubsec::observation_space}) and an action space that encompasses common operations for mobile UI navigation
(Section~\ref{subsubsec::action_space}).

\subsection{Toolkits for Multi-dimensional Evaluation}
\label{subsec::eval_dims}

\subsubsection{Trajectory-based Evaluator}
\benchmarkName{} uses trajectory-based evaluators to assure robust evaluation to end-to-end effectiveness.
In mobile apps, multiple ways to access the same functionality, known as alternative paths, are prevalent~\cite{lin2022route}. To mitigate the impact of these alternative paths on our evaluations, we employ a trajectory-based evaluation method~\cite{zhou2023webarena} and manually craft \textit{evaluators} to assess whether a trajectory generated by a model's navigation accomplishes the given task.
The evaluator functions as a boolean mechanism that assesses the success of navigation based on the model-generated trajectory $T$. Specifically, it checks whether $T$ satisfies the predefined criteria for task success by verifying the attributes and conditions at critical steps in the navigation process.  

A trajectory-based evaluator in \benchmarkName{} is defined by two key components: a \textit{list} and an \textit{order}. The \textit{list} contains elements that can either be \textit{basic assertions} or recursively defined \textit{sub-evaluators}. The \textit{order} specifies the evaluation sequence for the elements within the \textit{list}, which can be sequential, consecutive, or presence.

\benchmarkName{} defines two types of basic assertions to support the evaluator:
(1) \textit{StopPage} and \textit{LastAction} assertions, which focus on the final phase of UI navigation, such as the terminal UI screen or the last action executed before the stop action.
(2) \textit{FindAction}, \textit{FindElement}, and \textit{FindElementByAction} assertions, which target the presence of specific actions or elements during navigation.

By combining these basic assertions and sub-evaluators into a structured list, the trajectory-based evaluator enable robust evaluation of navigation trajectories under diverse conditions, providing a reliable assessment of the model's navigation capabilities.

Each task is assigned one or more evaluators. Upon generating a trajectory for a given task, the evaluators are invoked to assess the end-to-end effectiveness of the foundation model. To quantify performance, we employ two metrics: success rate (SR) and average completion proportion (ACP). SR represents the proportion of generated trajectories that successfully satisfy all evaluators, serving as an indicator of complete success. In contrast, ACP measures the average proportion of evaluators satisfied by a generated trajectory relative to the total number of evaluators. This metric offers a finer-grained perspective on partial success, capturing the degree to which the generated trajectory meets the requirements.

\subsubsection{Knowledge Probing}
To evaluate the goal understanding capability and app knowledge proficiency of foundation models, we adopt prompt-based knowledge probing techniques~\cite{vulic2020probing,alivanistos2022prompting}.
The knowledge probing process evaluate both the extent and accuracy of its understanding of mobile UI navigation by posing targeted questions.
Specifically, this approach utilizes multiple-choice questions (MCQs) and binary questions (BQs) to systematically probe the model’s knowledge (Section~\ref{subsubsec::fine_grained_knowledge_prob}).

\subsubsection{Completion Judgment}
Planning is a fundamental component in the domain of UI navigation, as it empowers models to efficiently achieve predefined goals. A critical aspect of planning involves the ability to accurately recognize task completion, which is essential for preventing errors such as prematurely halting or unnecessarily prolonging actions. To evaluate this capability, we employ the completion judgment task (Section~\ref{subsubsec::completion_task}), designed to assess the model's proficiency in determining whether a task has been successfully completed.

\subsubsection{Low-level Instruction}
Grounding capability plays a crucial role in enabling effective UI navigation, as it directly impacts the ability of models to associate low-level instructions with specific UI elements and actions. In the context of UI action grounding~\cite{li2020mapping,burns2022dataset,ran2024guardian} (Section~\ref{subsubsec::grounding_task}), the task requires a model to interpret an instruction (e.g., "click the search icon") and identify both the correct UI element and the corresponding action to execute. Grounding serves as a foundational skill that bridges perception and action, enabling robust and efficient UI navigation workflows.

\subsubsection{Invariants}

As described in Section~\ref{subsec::prompt}, even the baseline technique has multiple distinct instructions specifying the output formats and navigation rules (e.g., do not repeat erroneous actions), not to mention sophisticated agent designs~\cite{song2023llmplanner,yao2023tree}.
Analogous to using program invariants for verifying the correctness of programs~\cite{ernst1999dynamically}, we use invariants to evaluate the model capability of following specific instructions during mobile UI navigation.
Invariants are conditions or properties that should remain true during the navigation of a model on the mobile app. 
Violating such invariants will invalidate the mobile UI navigation.
Using invariants, \benchmarkName{} evaluates three key aspects of performance. 
\textbf{Repetition} measures the proportion of repeated actions generated during the end-to-end trace generation process, reflecting the model's ability to avoid redundant operations.
\textbf{Format error} assesses the proportion of incorrectly formatted actions produced during the end-to-end trace generation process, indicating the model's adherence to specified output formats.
\textbf{Focused context} (detailed in Section~\ref{subsubsec::fine_grained_invariant}) removes complex instructions to exclusively evaluate the model's capability to produce outputs in the required format without additional distractions.

\begin{table}[t]
\footnotesize
\centering
\begin{threeparttable}
\caption{Action Space of \benchmarkName{}.}
\label{table::action_space}
\begin{tabular}{cccc}
\toprule
\textbf{Action Type} & \textbf{Description}\\
\midrule
\texttt{click [elem]} & click the element \\
\texttt{longclick [elem]} & long click the element \\
\texttt{text [elem] [string]} & text the given string on the element\\
\texttt{swipe [elem] [dir]} & swipe the element in the given direction  \\
\midrule
\texttt{click [x,y]} & click (x,y) coordination on the screen\\
\texttt{longclick [x,y]} & long click (x,y) coordination on the screen\\
\texttt{text [x,y] [string]} & text the given string on the (x,y) coordination\\
\texttt{swipe [x1,y1] [x2,y2]} & swipe from (x1,y1) to (x2,y2) \\
\midrule
\texttt{press [back]} & Navigate to the previous screen\\
\texttt{press [home]} & Navigate to the Home screen  \\
\texttt{press [restart]} & Navigate to the home screen of the app  \\
\texttt{press [wait]} & Wait for page rendering and do nothing\\
\texttt{press [enter]} & Send the Enter event\\ 
\texttt{press [stop]} & Stop exploration and complete the task\\
\bottomrule
\end{tabular}
\end{threeparttable}
\vspace{-5pt} % Adjust this to keep the paper tight
\end{table}

\subsection{Task Collection}
\label{subsec::task_collection}

\subsubsection{Goal-based UI Navigation Task.}
\label{subsubsec::ui_navigation_task}

Each UI navigation task consists of a natural-language-based goal instruction $I$, describing the high-level goal of the task, and a reference trajectory $T =\{u_0, a_1, u_1,..., u_{n-1}, a_n\}$ (i.e., a UI transition sequence interleaved with UI screens and UI actions) confirming the task feasibility on the given app.
To comprehensively evaluate the effectiveness of existing models, we need to collect UI navigation tasks with different levels of difficulties.

\noindent\textbf{Testing task collection.}
One of the major applications of goal-based UI navigation is to automate UI test generation, which is notoriously time-consuming and labor-intensive.
To evaluate the effectiveness of existing models in performing goal-based UI navigation to meet real industry standards, we reuse in-house tests used daily to ensure the quality of \appX{}.
\appX{} is a highly popular mobile app with more than \textbf{one billion} monthly active users globally. 
Three quality assurance engineers with over three years of working experience use the test automation platform to run the test cases to assure their reproducibility without dependency on server-side configuration or account status.
We use the requirement document of the original test cases as the task instructions.
For test cases lacking requirement documents, the quality assurance engineers manually write high-level test descriptions as the task instructions.
Finally, we collect \taskWeChat{} testing tasks.

\noindent\textbf{User task collection.}
To cover as many popular functionalities of industrial apps with appropriate efforts, we first determine apps to collect tasks from and then determine tasks to collect.
In collaboration with professional developers and testing engineers from \companyX{}, we carefully identify \categoryTotal{} app categories suitable for automated testing and evaluation.
To ensure accessibility and ease of testing, we include only apps that either do not require login to access their main functionalities or allow login via Gmail accounts without CAPTCHA validation. Following this rigorous selection process, we finalize a set of \appTotal{} popular industrial apps for constructing \benchmarkName{}.
For each app, we identify its core functionalities and collect specific tasks corresponding to each functionality. For each task, we formulate a goal instruction $I$ that defines the high-level objective of utilizing the associated feature. We then interact with the app to record a reference trajectory that demonstrates how to successfully execute the task.
Finally, we collect \taskOpen{}{} user tasks from \appTotal{} highly popular apps cross \categoryTotal{} categories.

\begin{table}[t]
\footnotesize
\centering
\begin{threeparttable}
\caption{Models Benchmarked with \benchmarkName{}.}
\label{table::llm_information}
\begin{tabular}{lcccc}
\toprule
\textbf{Model} & \textbf{Creator} & \textbf{\# of Params} &\textbf{Modality} &\textbf{Open?}\\
\midrule
\textbf{GPT-4-Turbo\cite{openai_website}} & OpenAI & N/A & Text \& Image & \xmark \\
\textbf{GPT-4o\cite{hurst2024gpt}} & OpenAI & N/A & Text \& Image & \xmark \\
\textbf{GPT-4o-Mini\cite{openai_website}} & OpenAI & N/A & Text \& Image & \xmark \\
\textbf{Qwen-VL-Plus\cite{qwen_website}} & Alibaba & N/A & Text \& Image & \xmark \\
\textbf{Qwen-VL-Max\cite{qwen_website}} & Alibaba & N/A & Text \& Image & \xmark \\

\midrule
\textbf{DeepSeek-V2\cite{liu2024deepseek}} & DeepSeek & 21B/236B* & Text & \cmark \\
\textbf{Llama3-8B\cite{dubey2024llama}} & Meta AI & 8B & Text & \cmark \\
\textbf{Llama3-70B\cite{dubey2024llama}} & Meta AI & 70B & Text& \cmark \\
\textbf{Llama3.2-11B\cite{llama3.2}} & Meta AI & 11B & Text \& Image & \cmark \\
\bottomrule
\end{tabular}
\begin{tablenotes}
\small
\item Note: DeepSeek-V2 adopts the Mix of Expert (MoE) architecture, where only a subset of model parameters is activated.
\end{tablenotes}
\end{threeparttable}
\vspace{-10pt}
\end{table}

\subsubsection{Knowledge Probing Tasks.}\label{subsubsec::fine_grained_knowledge_prob}
To evaluate the goal-understanding capability and app knowledge proficiency of foundation models, we adapt prompt-based knowledge probing techniques~\cite{vulic2020probing,alivanistos2022prompting} by posing specific questions to the model to determine the extent and accuracy of its knowledge about mobile UI navigation.
We manually write a set of Multiple Choice Questions  (\textit{MCQs}) and Binary Questions (\textit{BQs}) to probe knowledge inside a model. 
An MCQ presents a question followed by several possible options, one of which is the correct answer.
An BQ is a question with only two possible answers "Yes" or "No".
Based on the answers to these MCQs/BQs, \benchmarkName{} can evaluate whether a model contains enough knowledge to understand goal instructions and know an app's functionalities and skills to achieve the tasks.
In total, \benchmarkName{} has 445 MCQs/BQs for evaluating the goal understanding capability and 445 MCQs/BQs for evaluating the knowledge proficiency of app functionalities.

\subsubsection{Completion Judgment Tasks}
\label{subsubsec::completion_task}

The completion judgment task is constructed based on the steps collected from the goal-based UI navigation tasks(Section \ref{subsubsec::ui_navigation_task}). In this task, the model is presented with the action history and the current UI screen for each step of a UI navigation task. The model must then decide whether to output ``continue'', indicating that the task should proceed, or ``stop'', signifying that the task has been completed. Specifically, the ground truth for the final step of each task is labeled as ``stop'', while the ground truth for all preceding steps is labeled as ``continue''.
In total, \benchmarkName{} contains 1190 steps labeled as ``continue'' and 244 steps labeled as ``stop'', adding up to a total of 1434 completion judgment tasks.

\begin{table*}[t]
        \centering
        \caption{End-to-end effectiveness of different models on \benchmarkName{}.}\label{table::main_results}
        \begin{threeparttable}
        \begin{tabular}{p{0.09\linewidth}p{0.17\linewidth}>{\centering\arraybackslash}p{0.09\linewidth}>{\centering\arraybackslash}p{0.10\linewidth}>{\centering\arraybackslash}p{0.09\linewidth}>{\centering\arraybackslash}p{0.10\linewidth}>{\centering\arraybackslash}p{0.09\linewidth}>{\centering\arraybackslash}p{0.10\linewidth}}
        \toprule
\multirow{2}{*}{\textbf{Modality}} & \multirow{2}{*}{\textbf{Model}} & \multicolumn{2}{c}{\textbf{User Tasks}} &\multicolumn{2}{c}{\textbf{Testing Tasks}} & \multicolumn{2}{c}{\textbf{All Tasks}} \\
    \cmidrule(lr){3-4} \cmidrule(lr){5-6} \cmidrule(lr){7-8}
     & & \textbf{SR} (\%) &  \textbf{ACP} (\%) & \textbf{SR} (\%) &  \textbf{ACP} (\%) & \textbf{SR} (\%) &  \textbf{ACP} (\%) \\
\midrule
\multirow{8}{*}{\textbf{Text}}
& \textbf{GPT-4-Turbo} & \textbf{31.1} & \textbf{34.5} & 0.0 & \textbf{6.8} & \textbf{16.6} & \textbf{21.5} \\
& \textbf{GPT-4o} & 28.7 & 33.5 & 0.0 & 6.5 & 15.3 & 20.8 \\
& \textbf{GPT-4o-Mini} & 25.0 & 29.4 & 0.0 & 5.7 & 13.3 & 18.3 \\
& \textbf{Qwen-VL-Plus}& 3.3 & 3.6 & 0.0 & 0.4 & 1.7 & 2.1 \\
& \textbf{Qwen-VL-Max} & 6.6 & 7.3 & 0.0 & 4.3 & 3.5 & 5.9 \\
& \textbf{DeepSeek-V2}  & 27.9 & 31.8 & 0.0 & \textbf{6.8} & 14.8 & 20.1 \\
& \textbf{Llama3-8B} & 2.5 & 2.7 & 0.0 & 0.0 & 1.3 & 1.5 \\
& \textbf{Llama3-70B} & 4.5 & 4.8 & 0.0 & 2.0 & 2.4 & 3.5 \\
\midrule
\multirow{5}{*}{\textbf{Vision}}
& \textbf{GPT-4-Turbo}  & 5.7 & 6.6 & 0.0 & \textbf{6.8} & 3.1 & 6.7 \\
& \textbf{GPT-4o} & \textbf{8.6} & \textbf{10.8} & 0.0 & 6.4 & \textbf{4.6} & \textbf{8.7} \\
& \textbf{GPT-4o-Mini} & 7.0 & 8.5 & 0.0 & 6.0 & 3.7 & 7.3 \\
& \textbf{Qwen-VL-Plus}& 1.6 & 1.9 & 0.0 & 0.1 & 0.9 & 1.1 \\
& \textbf{Qwen-VL-Max} & 3.7 & 4.7 & 0.0 & 4.1 & 2.0 & 4.4 \\
\bottomrule
\end{tabular}
% \begin{tablenotes}
% \small
 % \item \textit{Note:}
% \end{tablenotes}
\end{threeparttable}
% \vspace{-5pt}
\end{table*}

\subsubsection{Grounding Tasks.}
\label{subsubsec::grounding_task}
We extracted and deduplicated all single-step actions from the goal-based UI navigation tasks(Section \ref{subsubsec::ui_navigation_task}). For each action, we annotated the selected element in the screenshot with a bounding box and employed GPT-4o with a high temperature setting(0.75) to generate a concise, one-line natural language instruction describing the action.
This method resulted in the generation of 478 tasks. We then manually cleaned the data for all grounding tasks, removing 21 cases due to issues such as unclear screenshots and revising 37 GPT-4o-generated instructions to improve accuracy. Additionally, due to suboptimal annotations in the Goal-based UI Navigation Task dataset—such as ground truth bounding boxes targeting small text elements instead of the corresponding buttons—we corrected the ground truth bounding boxes for 74 tasks to ensure consistency and reliability.
Finally, \benchmarkName{} includes 457 grounding tasks with low-level instruction.

\subsubsection{Focused Context Tasks.}\label{subsubsec::fine_grained_invariant}
To better evaluate how effectively existing models can follow action format instructions, we design tasks that require generating responses in a strictly specified format while removing all other contextual instructions. These tasks emphasize prompting LLMs to produce outputs that adhere precisely to the given structure. For instance, a prompt might instruct the model to respond in a specific format such as: Respond with `input [123] [some text]'. Any additional instructions or extraneous context are excluded, ensuring the model's sole focus is on producing output that aligns with the prescribed format.
For this focused context task, we randomly generate 141 tasks using GPT-4o and manually check them to confirm that the generated tasks are accurate and adhere to the specified requirements.

\subsection{Benchmark Interface Design}\label{sec::benchmark::interface}

\lstset{
    basicstyle=\small\ttfamily,
    basewidth=4.5pt,
    breaklines,
    escapeinside={<@}{@>},
}

\subsubsection{Observation Space.}\label{subsubsec::observation_space}
\benchmarkName{} provides four types of observations for benchmarking language~\cite{brown2020language,codex}, vision~\cite{yuan2021florence}, and multi-modal models~\cite{alayrac2022flamingo,liu2024visual}.

\noindent\textbf{Image observation} presents the screenshot of the device as the observation of the current UI state.
Users primarily interact with mobile apps via screens, and image observation provides the most natural observation for foundation models~\cite{ran2022automated, koh2024visualwebarena, OSWorld, anthropic_computer_use}.

\noindent\textbf{Accessibility text observation} presents the UI content on the screen described by the accessibility API~\cite{AndroidAccessibility}.
The accessibility API creates a UI hierarchy tree, capturing all visible elements on the screen, including their states, properties, and contextual information. 
% Each element in the hierarchy is categorized either as a \texttt{View} or a \texttt{ViewGroup}, with additional metadata such as roles, labels, and actions that can be performed on them. 
These elements are structured hierarchically, mirroring the app's UI layout.
We faithfully describe the accessibility tree with plain text, using indents to represent the tree structure and unique indexes ``[i]" (where $i = 0,1..,$) to tag interactable UI elements.

\noindent\textbf{Simplified accessibility text observation}, following previous work~\cite{ran2023badge}, removes non-leaf UI elements and provides a list of descriptions of interactable UI elements in the UI hierarchy tree created by the accessibility API.

\noindent\textbf{Annotated image observation} integrates information from both the screenshot and the UI hierarchy tree by employing Set of Mark (SoM) prompting~\cite{yang2023setofmark}. We marks the center points of interactable UI elements—identified through analysis of the UI hierarchy tree—directly on the screenshot.

\subsubsection{Action Space.}\label{subsubsec::action_space}

Inspired by previous work on web UI navigation~\cite{yao2022webshop,zhou2023webarena}, we design an action space that emulates the touchscreen events and system navigation events commonly used in mobile UI navigation.
Table~\ref{table::action_space} presents the supported actions, categorized into three groups. 
The first and second groups include touchscreen operations, such as clicking, long-clicking, texting, and swiping.
Depending on the type of the observation space, \benchmarkName{} provides two types of touchscreen actions.
The first group uses element-grounded observation to execute touchscreen events associated with a specific UI element.
This ID is generated when traversing the UI hierarchy tree. With element IDs, the element selection is reduced to a grounding problem, thereby eliminating the agent's implementation efforts to map the agent's output to the real action.
The second group is designed for image observation, executing touchscreen events associated with specific coordinates on the screen. 
The third group of actions encompasses navigation-related events, including essential system navigation functions commonly used in mobile device interactions.

\section{Experiment Setup}\label{sec::setup}
\begin{table}[t]
\centering
\footnotesize
\caption{End-to-end effectiveness with different prompting strategies.}\label{table::rq1_ablation}
\begin{tabular}{p{0.20\linewidth}>{\centering\arraybackslash}p{0.13\linewidth}>{\centering\arraybackslash}p{0.13\linewidth}>{\centering\arraybackslash}p{0.13\linewidth}>{\centering\arraybackslash}p{0.13\linewidth}}
\toprule
\multirow{2}{*}{\textbf{LLMs}} & \multicolumn{2}{c}{\textbf{simplified accessibility tree}} & \multicolumn{2}{c}{\textbf{full accessibility tree}} \\
\cmidrule(r){2-3} \cmidrule(l){4-5}
& SR (\%) & ACP (\%)& SR (\%) & ACP (\%) \\
\midrule
\textbf{GPT-4o}   & 20.5 & 24.1 & 28.7 & 33.5 \\
\textbf{GPT-4o-Mini}   & 15.2 & 19.9 & 25.0 & 29.4 \\
\textbf{DeepSeek-V2}    & 19.7 & 23.3 & 27.9 & 31.8 \\
\textbf{Qwen-VL-Max}   & 3.7 & 4.9 & 6.6 & 7.3 \\
\midrule
\multirow{2}{*}{\textbf{VLMs}} & \multicolumn{2}{c}{\textbf{annotated image}} & \multicolumn{2}{c}{\textbf{image}} \\
\cmidrule(r){2-3} \cmidrule(l){4-5}
& SR (\%) & ACP (\%)& SR (\%) & ACP (\%) \\
\midrule
\textbf{GPT-4o}   & 8.6 & 10.8 & 5.7 & 6.2 \\
\textbf{GPT-4o-Mini}     & 7.0 & 8.5 & 3.3 & 3.6 \\
\textbf{Qwen-VL-Max}    & 3.7 & 4.7 & 2.5 & 2.9 \\
\bottomrule
\end{tabular}
\vspace{-5pt}
\end{table}

With \benchmarkName{}, our experiments aim to answer the following research questions:
\begin{itemize}
    \item \textbf{RQ1: End-to-end Effectiveness Analysis.} How effective are state-of-the-art foundation models on \benchmarkName{}?
    \item \textbf{RQ2: Multi-dimensional Analysis.} How effective are state-of-the-art foundation models in different dimensions?
    \item \textbf{RQ3: Agent Effectiveness Analysis.} How can the model's effectiveness affect the performance of UI navigation agent built on the model? 
\end{itemize}

\subsection{Foundation Models}
Table~\ref{table::llm_information} presents the information of \lmNum{} popular foundation models benchmarked and evaluated with \benchmarkName{}.
To comprehensively evaluate the state-of-the-practice of mobile UI navigation with foundation models, we experiment with open-source and proprietary models with different model sizes instead of using only proprietary models as previous works~\cite{zhou2023webarena, OSWorld}.
Due to space limits, we put the details of these models on our project website~\cite{sphinx-website}.

\subsection{Agent Implementation}\label{subsec::prompt}
We implement a ReAct~\cite{yao2022react} style agent for the initial benchmarking study.
We first provide a detailed overview of the mobile app environment, the description of action space, and domain knowledge that improves the effectiveness of UI navigation.
Then, we provide the model with the current observation, the task instruction, and the previously performed action.
We provide four types of observations detailed in Section~\ref{subsubsec::observation_space}
Based on these prompts, the model first performs chain-of-thought (CoT) ~\cite{wei2022chain} reasoning steps and outputs an action with the specified format (described in Section~\ref{subsubsec::action_space}).
The implementation of ReAct is unified across different foundation models, ensuring a fair comparison between them.
Due to space limits, we put detailed prompts on our project website~\cite{sphinx-website}.

\section{Experiment Results}\label{sec::results}

\subsection{RQ1: End-to-end Effectiveness}

Table~\ref{table::main_results} presents the main benchmark results of \lmNum{} models on \benchmarkName{}, from which we have four major observations.

\noindent\textbf{UI navigation is challenging for all FMs.}
Consistent with common belief on scaling law, larger models, such as GPT-4-Turbo and GPT-4o, achieves much higher SRs and ACPs compared with smaller models like Llama3-8B.
However, all benchmarked FMs achieves low effectiveness on \benchmarkName{}, with the highest SR being 16.6\% and ACP being 21.5\% across all tasks.
Notably, these FMs have specifically struggled with testing tasks, where no task is completed by any model and ACP tops at only 6.8\%.

\noindent\textbf{Testing tasks are more challenging than user tasks for existing models.}
As shown in Table~\ref{table::main_results}, none of existing FMs can successfully achieve one testing task on WeChat. 
After manual inspection with the assistance from QA engineers from WeChat, we find out three major reasons for their surprisingly low effectiveness.
First, testing tasks require more interaction steps compared with user tasks, leading to higher chances of failure. Testing tasks have 11.01 steps on average, while user tasks have 5.88 steps on average.
Second, actions in testing tasks are generally more context-sensitive than user tasks, with reduced tolerance of incorrect actions generated by LLMs.
For user tasks, it is possible to exploit alternative paths~\cite{lin2022route} to complete the task, while testing tasks require entering the functionality with the specified path.
Third, instructions for testing tasks tend to be complex and domain-specific, given that they are intended for professional testers.
On the contrary, user task instructions are expected to be easily understandable for average users, and we can expect LLMs to have fewer difficulties understanding them.
Since the testing tasks are written by QA engineers at \companyX{}, we conclude that existing FMs themselves may not be suitable for industrial mobile UI navigation alone without external assistance~\cite{ran2024guardian} or targeted fine-tuning.

\begin{table}[t]
\centering
\caption{Knowledge Probing.}\label{table::rq2_knowledge_probing}
\begin{tabular}{p{0.25\linewidth}>{\centering\arraybackslash}p{0.14\linewidth}>{\centering\arraybackslash}p{0.14\linewidth}>{\centering\arraybackslash}p{0.12\linewidth}>{\centering\arraybackslash}p{0.12\linewidth}}
\toprule
\multirow{2}{*}{\textbf{Models}} & \multicolumn{2}{c}{\textbf{Goal Understanding}} & \multicolumn{2}{c}{\textbf{App Knowledge}} \\
\cmidrule(r){2-3} \cmidrule(l){4-5}
& Original & Repaired & Original & Repaired \\
\midrule
\textbf{GPT-4-Turbo}     & 87.9  & 89.4& 95.5 & 95.5  \\
\textbf{GPT-4o}    & \textbf{91.7}  & \textbf{91.9} & \textbf{96.2} & \textbf{96.2}\\
\textbf{GPT-4o-Mini}      & \textbf{91.7}  & 91.7& 94.8 & 94.8  \\
\textbf{Qwen-VL-Plus}      & 40.4 & 86.3 & 64.7 & 90.1  \\
\textbf{Qwen-VL-Max}      & 57.5  & 85.6  & 85.8&  92.6\\
\textbf{DeepSeek-V2}     & 89.9  & 89.9 & 94.4 & 94.4\\
\textbf{Llama3-8B}      & 84.7 & 84.7 & 90.1 & 90.1 \\
\textbf{Llama3-70B}      & 89.7& 89.7& 95.3   & 95.3 \\
\textbf{Llama3.2-11B} & 86.7 & 86.7 & 94.4 & 94.4 \\ 

\bottomrule
\end{tabular}
\vspace{-20pt}
\end{table}

\begin{table*}[t]
\centering
\caption{Planning Capability.}\label{table::rq2_completion}
\begin{threeparttable}
\begin{tabular}{p{0.20\linewidth}>{\centering\arraybackslash}p{0.09\linewidth}>{\centering\arraybackslash}p{0.09\linewidth}>{\centering\arraybackslash}p{0.09\linewidth}>{\centering\arraybackslash}p{0.09\linewidth}>{\centering\arraybackslash}p{0.09\linewidth}>{\centering\arraybackslash}p{0.09\linewidth}}
\toprule

\multirow{2}{*}{\textbf{Models}} & \multicolumn{3}{c}{\textbf{Text Modality Accuracy(\%)}} & \multicolumn{3}{c}{\textbf{Vision Modality Accuracy(\%)}} \\
\cmidrule(r){2-4} \cmidrule(l){5-7}
& \textit{Continue.} & \textit{Stop.} & \textit{Perfect.} & \textit{Continue.} & \textit{Stop.} & \textit{Perfect.} \\

\midrule
\textbf{GPT-4-Turbo} & 98.5 & 47.1 & 41.8 & 99.8 & 18.4 & 18.4 \\
\textbf{GPT-4o} & 96.3 & 70.5 & \textbf{57.8} & 93.5 & 75.4 & \textbf{53.3} \\
\textbf{GPT-4o-Mini} & 98.6 & 32.0 & 27.9 & 98.0 & 27.9 & 24.2\\
\textbf{Qwen-VL-Plus} &  80.2 & 22.5 & 4.5 & 9.2 & 96.7 & 0.8 \\
\textbf{Qwen-VL-Max} & 92.4 & 36.9 & 24.6 & 90.3 & 52.1 & 30.3\\
\textbf{Llama3.2-11B} & 99.1 & 4.5 & 4.5 & 88.6 & 36.5 & 19.3\\
\textbf{Deepseek-V2} & 94.7 & 61.5 & 46.7 & N/A & N/A & N/A \\
\textbf{Llama3-8B}  & 96.1 & 24.6 & 18.0 & N/A & N/A & N/A \\
\textbf{Llama3-70B}  & 95.0 & 62.7 & 47.5 & N/A & N/A & N/A \\
\midrule
\textbf{Average}  & 94.5 & 40.3 & 30.4 & 79.9 & 51.2 & 24.4 \\
\bottomrule
\end{tabular}
\begin{tablenotes}
\small
 \item Note:   ``\textit{Continue.}'' and ``\textit{Stop.}'' are defined in Section~\ref{subsubsec::completion_task}. ``\textit{Perfect.}'' measures the success rate of a FM on all UI navigation tasks. a UI navigation task is counted as success if the FM succeeds on all completion judgment tasks originated from the UI navigation task.
\end{tablenotes}
\end{threeparttable}
\vspace{-5pt}
\end{table*}

\noindent\textbf{Language-based models outperform multi-modal models.}
Compared to language-based FMs, the multi-modal FMs achieve much lower effectiveness with UI screenshots as inputs, with all SRs and most ACPs less than 10\%.
To investigate whether the performance gap between LLMs and VLMs are impacted by prompt designs, we use GPT-3.5, GPT-4o, and DeepSeek-V2 to evaluate the impact of aggregating textual information in the UI hierarchy and use Qwen-VL-Plus, Qwen-VL-Max, and GPT-4o to evaluate the impact of using the Set of Mark (SoM) prompting strategy on vision-language models with the four types of observations (detailed in Section~\ref{sec::benchmark::interface}).
As shown in Table~\ref{table::rq1_ablation}, LLMs that take accessibility tree hierarchies as input achieve up to 28.7\% SR and 33.5\% ACP, while VLMs achieve up to only 8.6\% SR and 10.8\% ACP.

The primary reason for the low effectiveness of vision-language models is the visual grounding issues~\cite{zhou2023webarena,OSWorld}.
As shown in Table~\ref{table::rq1_ablation}, when FMs take images annotated with information from UI hierarchies as inputs, there are higher chances for them to successfully complete tasks compared with using only plain images.
The issue is further confirmed with our grounding evaluation provided by \benchmarkName{}, detailed in Sec~\ref{sec::eval::multi_dim::grounding}.
In addition, the accessibility tree provides more detailed information (e.g., types of UI elements, texts) about the UI than what are visible in the screenshots. 
As shown in Table~\ref{table::rq1_ablation}, compared to simplified accessibility tree describing only actionable UI elements, the full accessibility tree provides more information, which is helpful for FMs to make decisions.
While incorporating visual information could be beneficial, using UI element captioning~\cite{chen2020unblind} to transform visual information into textual descriptions~\cite{wu2024skill} can be a better way than directly prompting FMs with screenshots of the app.

\noindent\textbf{Dedicated benchmarks are required to evaluate the model performance on downstream tasks.}
While the series of Llama-3 models are shown competitive performance on widely used benchmarks such as MMLU~\cite{hendrycks2020measuring} and GPQA~\cite{rein2023gpqa},
they can hardly perform any UI navigation tasks on \benchmarkName{}, suggesting that results from generic benchmarks are not sufficient to reflect model performance in mobile UI navigation.
Thus, it is necessary to have dedicated benchmarks such as \benchmarkName{} for specific downstream tasks.

\vspace{3pt}
\noindent\fcolorbox{black}{gray!20}{
\begin{minipage}{\dimexpr 0.98\linewidth-2\fboxrule-2\fboxsep\relax}\Answer{RQ1}{
(1) State-of-the-art FMs still face significant challenges in mobile UI navigation, particularly in UI testing scenarios, indicating a substantial gap between current FM capabilities and the requirements of practical UI navigation tasks.
(2) Dedicated benchmarks are necessary for LLM performance evaluation. Generic benchmarks may not adequately capture the unique challenges and requirements of specific domains like mobile UI navigation, highlighting the importance of specialized evaluation frameworks.
(3) Despite the inherently visual nature of GUIs, language-based FMs currently show more promise than vision-based FMs,suggesting that the design of FM-based UI navigation agents should give priority to text modality being fed to FMs, while leveraging vision input primarily as a supplementary. 
}
\end{minipage}
}

\subsection{RQ2: Multi-dimensional Evaluations}

\subsubsection{Goal-understanding Capability.}
Table~\ref{table::rq2_knowledge_probing} presents the evaluation results of the goal-understanding capability. 
To ensure accurate assessment of the goal-understanding capability, we utilized DeepSeek-V2 to repair wrongly formatted outputs, which is also revealed as instruction following defects by \benchmarkName{} in Section~\ref{eval::rq2::inst}.
The results before and after the repair are labeled as ``Original'' and ``Repaired'', respectively.
From Table~\ref{table::rq2_knowledge_probing}, we have two major observations.

\noindent\textbf{Sufficient capability of goal-understanding.} 
Existing FMs exhibit remarkable effectiveness in goal-understanding tasks, with most achieving accuracy rates exceeding 85\%, and the best model GPT-4o reaching an impressive 91.9\%,  highlighting their strengths in understanding natural language instructions and intentions, which serves as a crucial foundation for effective UI navigation.

\noindent\textbf{Poor instruction following.}
While existing models have strong natural language understanding and goal comprehension, their power is hard to elicit even in simple question answering questions.
For example, Qwen-VL-Plus and Qwen-VL-Max cannot generate answers in the correct format, which is fatal in mobile UI navigation (detailed in Section~\ref{eval::rq2::inst}).
Consequently, eliciting the power of FMs in UI navigation remains a significant challenge.

\subsubsection{App Knowledge Proficiency.}
Table~\ref{table::rq2_knowledge_probing} presents the evaluation results of app knowledge proficiency.
Similar with the results of goal-understanding capability,  all models achieve over 90\% accuracy in app knowledge QA, demonstrating their capability to align user intentions with general app contents.
In addition, for some models, their limited ability to follow instructions substantially hinders their practical application in UI navigation tasks.

\subsubsection{Planning Capability.}\label{eval::rq2::plan}
Table~\ref{table::rq2_completion} presents the evaluation results of foundation models' planning capabilities in mobile UI navigation, specifically their ability to determine task completion status based on current progress. The evaluation focuses on models' judgment in deciding whether to continue or stop at each step of the navigation process. From Table~\ref{table::rq2_completion}, we have two major observations.

\noindent\textbf{Tendency to continue instead of stop.} Most FMs except for Qwen-VL-Plus tend to continue exploration instead of stop. In text modality, the accuracy of continuity judgment is 94.5\% on average, substantially higher than the accuracy of stop judgment being 40.3\% on average.
While the tendency to continue is alleviated for multi-modal FMs when fed with the vision modality, their planning capability is far from the expectations for industrial applications. 
Specifically, when the FM cannot timely stop the navigation process, it may execute unnecessary actions that invalidate the previously completed task, waste computational resources, and increase response latency through redundant operations.

\noindent\textbf{Accumulated planning error substantially decreases success rate.} The biases toward continuation or early stopping significantly impact the overall planning accuracy.
Taking GPT-4o with text modality as an example, even with 96.3\% and 70.5\% accuracy in ``Continue.'' and ``Stop.'' scenarios respectively, the overall success rate of perfectly completing an entire UI navigation task is only 57.8\%.

\subsubsection{Grounding Capability.}\label{sec::eval::multi_dim::grounding}
\begin{table}[t]
\centering
\caption{Grounding Capability.}\label{table::rq2_grounding}
\begin{threeparttable}
\begin{tabular}{p{0.25\linewidth}>{\centering\arraybackslash}p{0.12\linewidth}>{\centering\arraybackslash}p{0.12\linewidth}>{\centering\arraybackslash}p{0.12\linewidth}>{\centering\arraybackslash}p{0.12\linewidth}}
\toprule

\multirow{2}{*}{\textbf{Models}} & \multicolumn{2}{c}{\textbf{Text}} & \multicolumn{2}{c}{\textbf{Vision}} \\
\cmidrule(r){2-3} \cmidrule(l){4-5}
& Original & Repaired & Original & Repaired \\

\midrule
\textbf{GPT-4-Turbo} & 82.7 & 85.1 & 44.6 & 47.0\\
\textbf{GPT-4o} & \textbf{87.5} & \textbf{87.5} & \textbf{65.0} & \textbf{65.0} \\
\textbf{GPT-4o-Mini} & 81.4 & 81.8 & 53.4 & 53.6 \\
\textbf{Qwen-VL-Plus} & 4.6 & 32.4 & 3.3 & 12.3 \\
\textbf{Qwen-VL-Max} & 53.0 & 65.6& 27.4 & 30.2 \\
\textbf{Llama3.2-11B} & 0.2 & 68.5 & 5.0 & 46.6 \\
\textbf{Deepseek-V2} & 81.0 & 84.0 & N/A & N/A\\
\textbf{Llama3-8B}  & 0.2 & 66.7 & N/A & N/A\\
\textbf{Llama3-70B}  & 26.0 & 73.7 & N/A & N/A\\
\bottomrule
\end{tabular}
\end{threeparttable}
\vspace{-10pt}
\end{table}
Table~\ref{table::rq2_grounding} presents the results of grounding capability. In the text modality, GPT-4o achieved the highest accuracy at 87.5\%. In the vision modality, GPT-4o-Vision outperformed other models with an accuracy of 65.0\%. 
Due to the formatting error, some models like Qwen-VL-Plus, Llama3-8B, Llama3.2-11B cannot perform meaningful UI navigation. To show the capability of these models, we adopt Deepseek-V2 to repair the bad-formatted output. The result is shown as ``Original'' and ``Repaired''.
After inspecting the results, we have three explanations for the failure cases of grounding. 
First, UI hierarchies often lack comprehensive visual information~\cite{chen2020unblind}, such as the inability to describe image content.
Second, UI hierarchies are frequently inaccurate; for instance, clickable tags for some interactive elements are not correctly set to ``True".
Finally, vision modality models struggle to ground visual information in the UI navigation scenario.

While GPT-4o achieved a notable 87.5\% accuracy in the text modality, this performance remains insufficient for UI navigation tasks, which typically require multi-step UI grounding. 
To address the identified challenges, several potential directions for improvement can be explored.
First, UI information enhancement techniques could be developed to repair Android UI hierarchies~\cite{xie2022psychologically} and incorporate richer visual information~\cite{chen2020unblind,wu2024skill}, thereby improving UI grounding accuracy.
Second, leveraging both textual and visual inputs could help mitigate the limitations inherent in single-modality processing. Foundation models should also improve multi-modal integration capabilities.
Finally, fine-tuning foundation models on UI-specific tasks, supported by UI grounding datasets~\cite{rawles2024androidinthewild}, can further align model capabilities with the demands of real-world navigation scenarios.

\subsubsection{Instruction Following Capability.}\label{eval::rq2::inst}

Table~\ref{table::rq2_instruction_following} presents the percentages of three kinds of violations of existing FMs.
Results show that it is challenging for models to fully follow our instructions in mobile UI navigation scenarios: all models violate our repeated action requirement for about half of the time, and some models frequently generate malformed outputs.
In the focused context, violations against action format are significantly reduced for most models, where the best model GPT-4o achieves a zero violation rate.
It should be noted that, while the instruction following capabilities can be satisfactory in controlled environments, their performance drop in complex scenarios makes FMs less reliable in practical mobile UI navigation tasks.

To further enhance instruction-following capabilities, future work can focus on integrating instruction-following tasks from complex real-world scenarios into training datasets to improve agent reliability.
Another promising direction is to explore structured output generation techniques to ensure models strictly adhere to predefined formats.
Additionally, introducing external error detection and recovery mechanisms can help agent automatically identify and correct instances where the model fails to follow instructions~\cite{ran2024guardian}.

\begin{table}[t]
\centering
% \footnotesize
\caption{Failure rate of instruction following.}
\label{table::rq2_instruction_following}
\begin{threeparttable}
\begin{tabular}{p{0.25\linewidth}>{\centering\arraybackslash}p{0.18\linewidth}>{\centering\arraybackslash}p{0.20\linewidth}>{\centering\arraybackslash}p{0.20\linewidth}}
\toprule
\multirow{2}{*}{\textbf{Models}} & \multirow{2}{*}{\textbf{Repetition}} & \textbf{Format} & \textbf{Focused} \\
 & & \textbf{Error} & \textbf{Context} \\
\midrule
\textbf{GPT-4-Turbo}     & \textbf{49.6}  & 2.4 & 0.7 \\
\textbf{GPT-4o}    & 50.5  & \textbf{0.5} & \textbf{0.0} \\
\textbf{GPT-4o-Mini}      & 50.9  & 0.7 & \textbf{0.0}  \\
\textbf{Qwen-VL-Plus}      & N/A & 82.8 & 77.3\\
\textbf{Qwen-VL-Max}      & 58.0  & 11.9  & 73.8\\
\textbf{DeepSeek-V2}     & 50.1  & 1.5 & \textbf{0.0} \\
\textbf{Llama3-8B}      & N/A & 97.6 & 31.2\\
\textbf{Llama3-70B}      & 56.8 & 33.0 & 100.0\\
\bottomrule
\end{tabular}
\begin{tablenotes}
\small
 \item Note: The units are expressed in percentages (\%). \textbf{N/A} indicates meaningless UI navigation due to format error.
\end{tablenotes}
\end{threeparttable}
\vspace{-10pt}
\end{table}

\vspace{3pt}
\noindent\fcolorbox{black}{gray!20}{
\begin{minipage}{\dimexpr\linewidth-2\fboxrule-2\fboxsep\relax}\Answer{RQ2}{
(1) FMs demonstrate strong capabilities in natural language understanding and common knowledge reasoning required for UI navigation. However, they exhibit significant gaps in UI-specific capabilities such as UI grounding and precise task planning. These limitations suggest that while FMs provide a solid foundation for general comprehension, targeted fine-tuning focusing on UI-specific patterns and interactions is necessary for improved performance.
(2) In complex UI navigation scenarios, models face substantial challenges in instruction following and precise planning.
These limitations indicate that successful deployment of FMs in UI navigation may require either external support systems to handle structured output generation and validation or specialized fine-tuning focusing on instruction adherence and planning precision. 
}
\end{minipage}
}

\subsection{RQ3: Impact of Model Deficiency on UI Navigation Agents}

We further experiment with AppAgent~\cite{yang2023appagent}, a popular multimodal UI navigation agent designed to interact with smartphone apps, leverages automated exploration and foundation model reflection to construct app documents for enhanced UI navigation capabilities.

We conduct an end-to-end evaluation on a subset of \benchmarkName{} comprising 163 non-login tasks, considering token consumption.
We reuse the open-source implementation of AppAgent, allowing 10-minute autonomous exploration periods per app for document generation.
Due to the high costs associated with GPT-4-Turbo and AppAgent's autonomous exploration, GPT-4-Turbo is evaluated exclusively without document support.
Table~\ref{table::rq3_overall} presents AppAgent's performance on \benchmarkName{}. 
Even in its optimal configuration using GPT-4o, AppAgent achieved a mere 8.0\% SR without documents, which is comparable to ReAct's vision-based performance.
The incorporation of documents did not yield a significant improvement in success rates.
Notably, both configurations performed substantially below ReAct's text-modality results.

To further understand these challenges and investigate why AppAgent performs poorly on \benchmarkName{}, we conducted a case study analyzing its failed tasks on representative tasks from \benchmarkName{}.
Our case study reveals three major root causes accounting for the failure of AppAgent and all these root causes are revealed by the multi-dimensional evaluation of \benchmarkName{}. 

\noindent\textbf{Grounding defects impede the agent's ability to map concrete plans to UI elements.}
In a tip calculation task, while interacting with the Tip Calculator app, the agent was required to click ``Continue'' to grant authorization. Although AppAgent correctly summarizes its intent by stating, ``I have observed the permission request for the Tip Calculator app and am now tapping the `Continue' button to proceed with the task,'' it fails to correctly identify the appropriate ``Continue'' element based on its observations. Instead, it mistakenly clicked on an adjacent blank area, causing the task to be blocked and halting further progress.
This failure highlights the model's limited grounding capabilities especially when it takes images as input, as revealed in Section~\ref{sec::eval::multi_dim::grounding}

\begin{table}[t]
\centering
\caption{End-to-end success rate of UI navigation agents.}\label{table::rq3_overall}
\begin{tabular}{p{0.19\linewidth}>{\centering\arraybackslash}p{0.20\linewidth}>{\centering\arraybackslash}p{0.22\linewidth}>{\centering\arraybackslash}p{0.22\linewidth}}
\toprule
\multirow{2}{*}{\textbf{Agents}} & \multicolumn{3}{c}{\textbf{Models}} \\
\cmidrule(){2-4} 
& \textbf{GPT-4o} & 
\textbf{GPT-4o-Mini} & 
\textbf{GPT-4-Turbo} \\
\midrule
\textbf{ReAct} \\
\cmidrule(){1-1} 
\textbf{text}     & \textbf{32.5}  & \textbf{27.0} & \textbf{34.4} \\
\textbf{vision}   & 8.6           & 6.7           & 4.9 \\
\midrule
\textbf{AppAgent} \\
\cmidrule(){1-1}
\textbf{w/o docs}        & 8.0           & 3.7           & 5.5 \\
\textbf{w/ docs}        & 11.0           & 8.0           & / \\
\bottomrule
\end{tabular}
\vspace{-15pt}
\end{table}

\noindent\textbf{Insufficient planning results in unnecessary and erroneous actions.}
When performing the ``View Politics category'' task in the ABC News app using GPT-4-Turbo as the foundation model, AppAgent initially navigates to the politics category through a series of effective actions. However, it subsequently clicks on a news article, introducing an unnecessary step that caused the task---already completed at that point---to be marked as incorrect. This behavior underscores the model's lack of precise task planning capability, which leads to redundant or erroneous actions that compromise success rates as revealed in Section~\ref{eval::rq2::plan}.

\noindent\textbf{Instruction-following errors disrupt UI navigation.}
In many cases, AppAgent struggles to execute tasks due to difficulties in parsing its own outputs.
The generated actions frequently included extraneous characters, leading to parsing failures and ultimately causing the entire UI navigation task to stall.
While the prompt design and workflow of AppAgent are more complex than those of ReAct, the foundational model's instruction-following capabilities were not robustly adhered to. This increased complexity often resulted in formatting errors, as revealed in Section~\ref{eval::rq2::inst}.
Instead of relying on FM's own instruction following capability, adopting an external system for enforcing instruction following~\cite{ran2024guardian} can be a more practical approach.

\vspace{3pt}
\noindent\fcolorbox{black}{gray!20}{
\begin{minipage}{\dimexpr 0.98\linewidth-2\fboxrule-2\fboxsep\relax}\Answer{RQ3}{(1) Fundamental deficiencies in FMs significantly limit the effectiveness of UI navigation agents built upon them. Even sophisticated agents struggle to overcome the underlying model's limitations in grounding, planning, and instruction following.
(2) Multi-dimensional evaluation with \benchmarkName{} should precede agent design. \benchmarkName{} allows developers to identify specific weaknesses in model capabilities, make informed decisions about model selection, design targeted mitigation strategies, and focus development efforts on areas where the FM needs the most support. These findings emphasize the importance of understanding and addressing FM limitations at their source, rather than attempting to compensate for them solely through agent design.
}
\end{minipage}
}

\section{Threats to Validity}\label{sec::threats}
The primary internal threat is the degree to which the models used in our experiments are representative of true practice.
To mitigate this, we selected recent models that have achieved SOTA performance on popular benchmarks at their time of proposal, and all models come from established industry sources.
The primary external threat is whether tasks collected in \benchmarkName{} are representative of real-world scenarios. We addressed this through two approaches: (1) collecting common user tasks from highly popular apps used by billions of users, and (2) incorporating real test cases from WeChat's quality assurance process, ensuring our benchmark reflects both general user behaviors and industrial testing requirements.
\section{Conclusion}\label{sec::conclusion}
In this paper, we have presented \benchmarkName{}, a multi-dimensional benchmark for evaluating the capabilities of foundation models (FMs) in mobile UI navigation tasks. \benchmarkName{} distinguishes itself with two key contributions: (1) a multi-dimensional assessment framework examining grounding, planning, and instruction-following capabilities, and (2) a diverse collection of real-world UI navigation tasks from industrial applications and internal test cases at WeChat.
Evaluations of \lmNum{} FMs reveal that while FMs demonstrate strong capabilities in natural language processing capabilities and common knowledge, they face significant challenges in UI-specific capabilities and instruction following in UI navigation scenarios. 
These fundamental limitations cascade into substantial performance degradation when the FMs are integrated into UI navigation agents, highlighting the importance of addressing model deficiencies at their source or with external system supports rather than relying solely on agent architecture improvements.
\benchmarkName{} enables developers to identify and target specific weaknesses in FMs before deployment in UI navigation systems, paving the way for more focused improvements in model capabilities and more effective UI navigation solutions.

\balance
\bibliographystyle{ACM-Reference-Format}
\bibliography{ref}

\end{document}